\documentclass[11pt]{article}

\usepackage[utf8]{inputenc}
\usepackage[margin=1in]{geometry}
\usepackage{amsmath}
\usepackage{graphicx}
\graphicspath{{./}{./figures/}}

\usepackage{hyperref}

\title{The Value of Internal Memory for Population Growth in Varying Environments}

\author{Leo Law, BingKan Xue}

\date{Department of Physics, University of Florida, Gainesville, FL 32611, USA}

\begin{document}

\maketitle

\begin{abstract}
In varying environments it is beneficial for organisms to utilize available cues to infer the conditions they may encounter and express potentially favorable traits. However, external cues can be unreliable or too costly to use. We consider an alternative strategy where organisms exploit internal sources of information. Even without sensing environmental cues, their internal states may become correlated with the environment as a result of selection, which then form a memory that helps predict future conditions. To demonstrate the adaptive value of such internal memory in varying environments, we revisit the classic example of seed dormancy in annual plants. Previous studies have considered the germination fraction of seeds and its dependence on environmental cues. In contrast, we consider a model of germination fraction that depends on the seed age, which is an internal state that can serve as a memory. We show that, if the environmental variation has temporal structure, then age-dependent germination fractions will allow the population to have an increased long-term growth rate. The more organisms can remember through their internal states, the higher growth rate a population can potentially achieve. Our results suggest experimental ways to infer internal memory and its benefit for adaptation in varying environments.
\end{abstract}

\newpage
\section{Introduction}

Organisms can adapt to a varying environment by diversifying their traits among individuals of the same population. A common form of such diversity is dormancy, where some individuals enter a dormant state while others remain active \cite{Baskin2014, Lennon2011, Lennon2021}. Those that are active will contribute to the growth of the population under good environmental conditions, but will be vulnerable to periods of harsh conditions. On the other hand, the dormant individuals are often tolerant to environmental stress and thus help preserve the population during harsh periods. For example, in a bacterial population, while most cells grow and divide normally, some cells randomly switch to a reversible dormant state called persister cells, which makes them tolerant to antibiotics when normal cells would perish \cite{Balaban2004, Harms2016, Manuse2021}. Other examples include seed dormancy in plants, dauer larva in nematodes, diapause in insects, etc. \cite{Baskin2014, Lennon2021, Simon2011, Grimbergen2015}. These are thought to be a strategy known as diversified bet-hedging \cite{Seger1987, Philippi1989}, in which organisms express different traits with some probability to create diversity in the population, so as to increase the long-term growth rate of the population under environmental variations \cite{Cohen1966, Kussell2005, Donaldson-Matasci2008}.

In the simplest form, bet-hedging organisms have fixed probabilities of expressing different traits \cite{Cohen1966}. But more generally, organisms can sense cues from the environment that will influence these probabilities \cite{Cohen1967, Clauss2000}. Such cues may be indicative of future environmental conditions, so that the organisms may bias the probabilities towards traits that are favorable in the likely environment. It has been shown that the information contained in the cue about the environment will contribute to an increase in the population growth rate \cite{Cohen1967, Donaldson-Matasci2010, Rivoire2011}. However, sensing and responding to environmental cues may come at a cost, as it requires the expression of specific sensors and signaling mechanisms \cite{Auld2010}. Besides, there may not be enough time for the organisms to respond to the cues through phenotypic plasticity, as the environment may have changed by the time the trait is developed \cite{DeWitt1998, Murren2015}. Therefore, it is not always beneficial to rely on environmental cues.

Besides external signals, the behavior of organisms can be influenced by their internal states, such as physiological or metabolic states \cite{Higginson2018}. One example is the reserve level -- a starved animal may choose to forage more aggressively despite higher predation risk \cite{McNamara1987, Higginson2018}. Another example is the age of the organism -- it is known that the age of seeds can affect germination in annual plants \cite{Philippi1993}. These internal states are not sensors that directly measure the external environment. However, they may become correlated with the environment as a result of selection, because certain states are associated with higher fitness in past environmental conditions and thus become more common in the population. Therefore, the distribution of such internal states among the population can potentially provide information about the environment, which may be utilized by the organisms.

We will study an example of this situation and show that internal states of the organisms can indeed serve as internal cues to help them adapt to varying environmental conditions. Such internal states effectively provide a memory about the past outcomes of selection, which helps predict the future environment. Moreover, we show that a larger memory capacity enables higher gains in the population growth rate. Our results suggest that internal states that were not developed for sensing the environment could nevertheless be co-opted as internal cues for adaptation, which would save the cost of sensors and may thus be a more efficient strategy.

To study adaptation in varying environments, we will use seed dormancy as our main example. Seeds of annual plants will either germinate or stay dormant in a given year. While dormancy sacrifices the short-term fitness of the seeds, it preserves the population from a catastrophically bad year with very low yield, and thus results in higher long-term benefit. This has been studied as a classic model of bet-hedging \cite{Cohen1966, Cohen1967}, supported by the fact that dormant seeds eventually germinate under similar environmental conditions \cite{Philippi1993}, and that the germination fraction is negatively correlated with local environmental variability \cite{Venable2007}. It is known that germination is influenced by environmental cues, such as temperature, humidity, and the number density of surrounding seeds \cite{Clauss2000, Gremer2014}. Moreover, there is evidence that the probability a seed will germinate also changes with the age \cite{Kalisz1991, Kalisz1992, Philippi1993}. However, the adaptive value of such age dependence in germination has not been fully studied \cite{Valleriani2006, Lennon2021}. It was shown in \cite{Valleriani2006} that the evolutionarily stable probability of germination does not depend on seed age if there is no density dependence. Yet, their model did not include temporal correlation in the environmental variation, which is crucial for memory to be useful in predicting future environments \cite{Lambert2014, Marzen2018, Rescan2020}. We will show that, when there is temporal structure in the environmental variation, age-dependent germination probabilities can increase the long-term growth rate of the seed population.

\section{Background}

\subsection{Cohen's model of seed dormancy}

\begin{figure}
\centering
\makebox[\textwidth][c]{\includegraphics[width=7.0in]{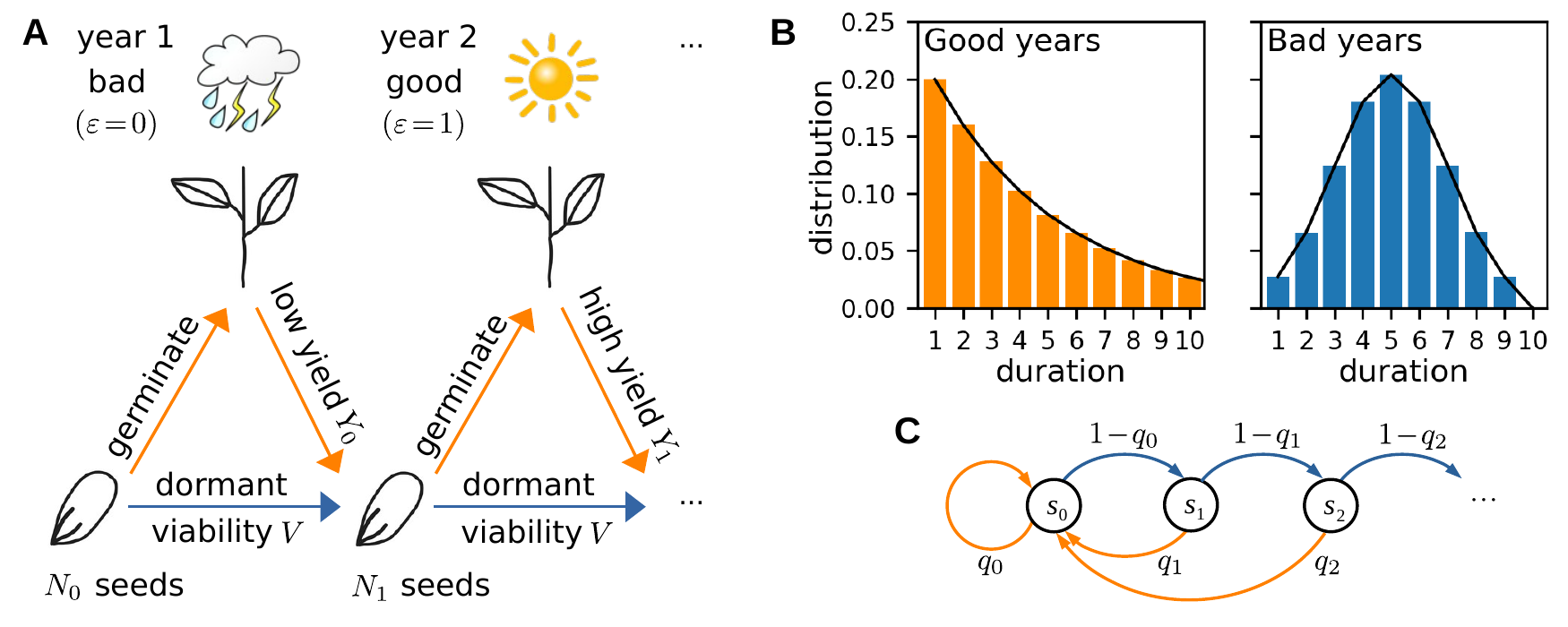}}
\caption{\small (A) Schematic illustration of Cohen's model of seed dormancy in annual plants. Each year may be good or bad for plant growth. A seed can either germinate to produce a yield $Y_\varepsilon$ that depends on the environmental condition $\varepsilon$, or stay dormant with a probability $V$ of still being viable next year. The number of seeds at the end of year $t$ is $N_t$. The parameter values used in our calculations are $Y_0 = 0$, $Y_1 = 4$, $V = 0.9$. (B) The distribution of duration of consecutive good years and bad years. We choose the duration of good years to follow a geometric distribution with a mean of $5$, and the duration of bad years to have a Gaussian distribution with a mean and standard deviation of $5 \pm 2$ cut off at 0 and 10. (C) A state diagram that represents the seed age. Each state $s_\alpha$ represents a seed of age $\alpha$. Blue arrows represent dormancy that increases the age by $1$; orange arrows represent germination that may produce new seeds of age $0$. Weights on the arrows represent the probability of germination or dormancy.}
\label{fig:durations}
\end{figure}

Let us first briefly review the idea of bet-hedging and how information emerges as a central quantity in determining the long-term growth rate of the population. We will follow the classic model of seed dormancy in annual plants by Cohen \cite{Cohen1966, Cohen1967}, as illustrated in Fig.~\ref{fig:durations}A. Each year can be ``good'' (denoted as environment $\varepsilon = 1$) or ``bad'' ($\varepsilon = 0$) for the plant. Seeds that germinate (``phenotype'' $\phi = 1$) in a good year will be able to grow and produce a large number ($Y_1$) of new seeds. However, in a bad year, germinated plants will have a low yield ($Y_0$). We will set $Y_0 = 0$ and denote $Y_1 = Y$ for simplicity, meaning that germinating in a bad year will result in no offspring. All germinated plants perish at the end of the year, regardless of their yield. Seeds that stay dormant ($\phi = 0$) will remain viable the next year with probability $V$. Thus, the fitness of a seed in a given environment can be summarized by the matrix $f_{\varepsilon\phi} = \left( \begin{smallmatrix} V & 0 \\ V & Y \end{smallmatrix} \right)$. In addition, we assume that the number of consecutive good years follows a geometric distribution, whereas that of bad years has a narrow distribution (see Fig.~\ref{fig:durations}B and Appendix~\ref{sec:numerical}). This is meant to describe the scenario where good growth conditions are disrupted by random occurrence of disasters that affect growth for a characteristic number of years.

In the simplest case where seeds receive no environmental cues, the fraction of seeds that germinate each year is assumed to be a constant, denoted by $q$. In a good year, the total number of seeds will grow by a factor $(1-q) V + q Y$, whereas in a bad year, the number of seeds will reduce to only a fraction $(1-q) V$ of the previous year. The long-term growth rate of the population will be given by (see derivation in Appendix~\ref{sec:derivation})
\begin{equation} \label{eq:Cohen2}
\Lambda = p \log \big( (1-q) V + q Y \big) + (1-p) \log \big( (1-q) V \big) \;,
\end{equation}
where $p$ is the frequency of good years and $(1-p)$ is that for bad years. The germination fraction that maximizes the long-term growth rate is
\begin{equation} \label{eq:Cohensol}
q^* = \frac{p \, Y - V}{Y - V}
\end{equation}
for $p > V/Y$ and 0 otherwise. In the limit of high yield ($Y \gg V$), this leads to the classic result $q^* \approx p$, which means the optimal germination fraction should match the frequency of good years \cite{Cohen1966}. The model can be extended to seeds that receive some external cue ($\xi$) about the environment \cite{Cohen1967}. In this case, the optimal germination fraction will depend on the cue. As a result, the population can grow faster than without the cue (see Appendix~\ref{sec:derivation}).

\begin{figure}
\centering
\includegraphics[width=5.5in]{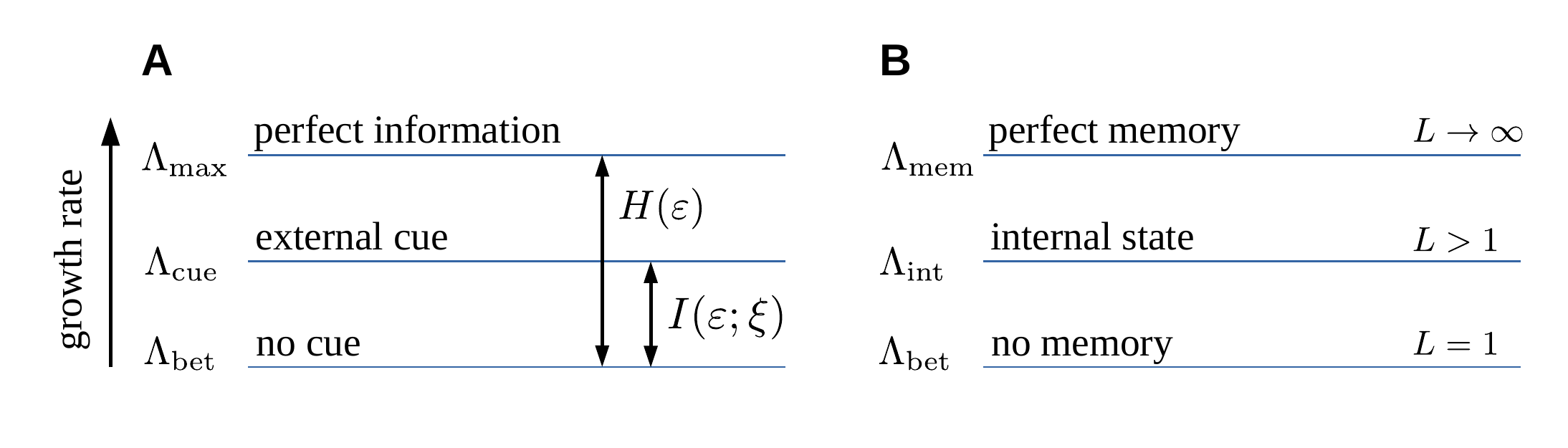}
\caption{\small The long-term growth rate $\Lambda$ of populations with different sources of information. (A) The value of external cues: $\Lambda_\textrm{max}$ is the maximum possible growth rate attainable if the population has perfect information about the future environment. $\Lambda_\textrm{bet}$ is the highest growth rate achievable by a bet-hedging population without receiving cues, which is suppressed by the entropy of the environment $H(\varepsilon)$. $\Lambda_\textrm{cue}$ is the growth rate when the population utilizes a cue $\xi$ that has a mutual information $I(\varepsilon;\xi)$ with the environment. (B) The value of internal memory: Organisms can utilize their internal states as memory, such that their behavior depends on which state they are in. $\Lambda_\textrm{bet}$ from bet-hedging also represents the case with no memory, which corresponds to having only one internal state ($L=1$). More states ($L > 1$) provides larger memory capacity and allows a higher growth rate $\Lambda_\textrm{int}$ for the population. $\Lambda_\textrm{mem}$ is the highest growth rate achievable by organisms with a perfect memory $(L\to\infty)$ of their lineage history.}
\label{fig:levels}
\end{figure}

These well-known results are summarized schematically in Fig.~\ref{fig:levels}A. At the top level is the maximum possible growth rate $\Lambda_\textrm{max}$, which is attainable only if individuals have perfect information about future environmental conditions and respond accordingly, i.e., germinate if it will be a good year and go dormant if it will be bad. On the other hand, if there is no environmental cue, the best strategy is bet-hedging with fixed probabilities, which achieves a growth rate $\Lambda_\textrm{bet}$. This is less than $\Lambda_\textrm{max}$ by an amount $H(\varepsilon)$, which is the Shannon entropy from information theory that quantifies the uncertainty of the varying environment (See Appendix~\ref{sec:derivation}). However, if a cue $\xi$ is used to help predict the environment, the population can increase the growth rate from $\Lambda_\textrm{bet}$ to $\Lambda_\textrm{cue}$, up by an amount $I(\varepsilon;\xi)$ that is equal to the mutual information between the cue and the environment (Appendix~\ref{sec:derivation}). Note that $\Lambda_\text{cue}$ is still not as high as $\Lambda_\textrm{max}$ unless the cue is fully accurate. The relations between these growth rates illustrated here (similar to plots in \cite{Donaldson-Matasci2010, Xue2018}) show that, in order for the population to better adapt to varying environments, it must utilize available sources of information about the environment.

\subsection{Internal source of information} \label{sec:internal}

Instead of sensing external cues, below we consider another possibility for organisms to use their internal states as a source of information. We will use the age of seeds as an example. The state diagram representing seed ages are illustrated in Fig.~\ref{fig:durations}C, where a state $s_\alpha$ represents a seed of age $\alpha$. A blue arrow represents a seed going into dormancy for one year, so that the age is increased by $1$. An orange arrow represents a seed that germinates and potentially produces new seeds, which will have age $0$. The weights on the arrows represent the probability of germination or dormancy. For a simple bet-hedging strategy without any cues, the probability of germination will be a constant, which equals $q^*$ from Eq.~(\ref{eq:Cohensol}), independent of the seed age. We will study the case where the germination fraction can depend on the seed age, and show that the population can acquire information from this internal state to achieve a higher growth rate.

\section{Results}

\subsection{Seed age as an internal cue}

\begin{figure} 
\centering
\includegraphics[width=3.4in]{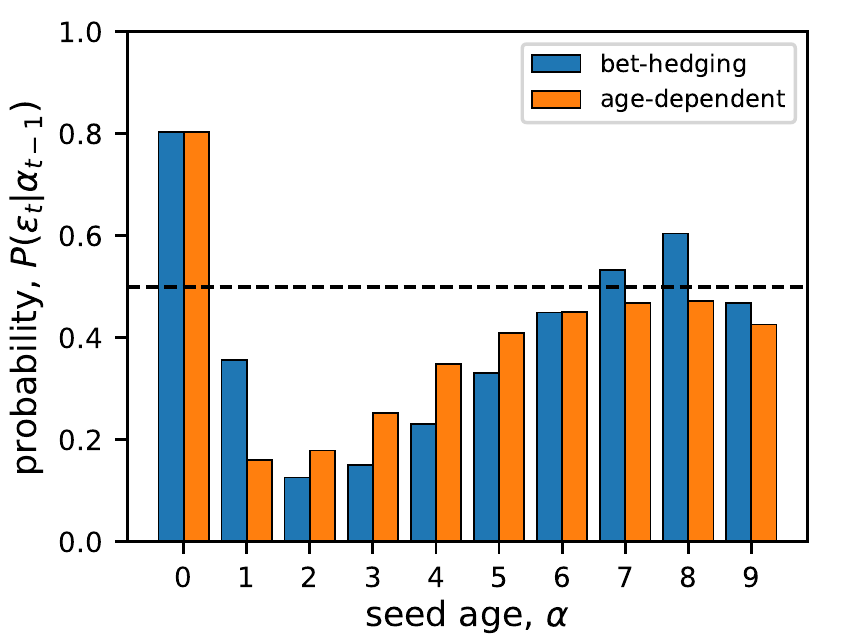}
\caption{\small Probability of the coming environment $\varepsilon_{t}$ conditioned on the seed age $\alpha_{t-1}$ at the beginning of year $t$, as calculated by simulating a lineage of seeds. Dashed line is the marginal probability of the environment, which would indicate that the seed age is uncorrelated with the environment. Blue bars are when the population uses a bet-hedging strategy with a constant germination fraction. Orange bars are when the germination fraction depends on the seed age to maximize population growth rate. In both cases the seed age is correlated with the environment and thus useful as an internal cue.}
\label{fig:correlation}
\end{figure}

We first study whether the seed age as an internal state contains useful information about the environment. Let $\alpha_{t-1}$ be the seed age at the beginning of year $t$, and $\varepsilon_{t}$ be the coming environment that year. If $\alpha_{t-1}$ has no information about the environment, then it will be statistically independent of $\varepsilon_{t}$, i.e., $P(\varepsilon_{t}|\alpha_{t-1}) = P(\varepsilon_{t})$. Therefore, whether seed age is informative about the environment can be inferred from the conditional probability $P(\varepsilon_{t}|\alpha_{t-1})$. To calculate that, we simulate a sufficiently long sequence of environments, denoted by $\varepsilon_t$ for each year $t$. We also simulate a single lineage of plants that uses the constant germination fraction $q^*$. Each year the seed can either germinate or stay dormant, and the probability of choosing the phenotype $\phi_t$ is further weighted by the fitness $f(\varepsilon_t,\phi_t)$ to account for selection (see procedure in Appendix~\ref{sec:lineage}). The seed age along the lineage is recorded as $\alpha_t$. From the sequences of $\varepsilon_t$ and $\alpha_t$, we estimate the joint probability distribution $P(\varepsilon_{t}, \alpha_{t-1})$, from which the conditional probability $P(\varepsilon_{t}|\alpha_{t-1})$ is calculated. As shown in Fig.~\ref{fig:correlation}, the probability of the environment $\varepsilon_{t}$ does depend on the seed age $\alpha_{t-1}$. This means that knowing the seed age allows a more accurate prediction of the coming environment. Therefore, it is possible for the population to ``co-opt'' the seed age as an ``internal cue'' for the environment. In analogy to the case of external cues, we expect that such information can be used to increase the long-term population growth rate.

We therefore consider a strategy where the germination fraction depends on the seed age, denoted by $q_\alpha$ and represented by weights on the arrows in Fig.~\ref{fig:durations}C. To calculate the long-term growth rate, let $\boldsymbol{N}$ be a vector that represents the age-structured population, with components $N_\alpha$ being the number of seeds of age $\alpha$. The dynamics of $\boldsymbol{N}$ is described by a matrix $\boldsymbol{M}(\varepsilon;\boldsymbol{q})$ that depends on the environment $\varepsilon$ and the germination fractions $\boldsymbol{q}$ (with components $q_\alpha$),
\begin{equation} \label{eq:Mat}
    \boldsymbol{M}(\varepsilon;\boldsymbol{q}) =
    \left( \begin{array}{ccc}
    q_0 \, Y_\varepsilon & \!\!\! q_0 \, Y_\varepsilon & \cdots \\[4pt]
    (1\!-\!q_1) V & \!\!\! 0 & \cdots \\
    0 & \!\!\! (1\!-\!q_2) V & \ddots  \\
    \vdots & \!\!\! \ddots & \ddots    
    \end{array} \right)
\end{equation}
Each year, the population vector is multiplied by the matrix that corresponds to the current environment $\varepsilon_t$,
\begin{equation} \label{eq:vecN}
\boldsymbol{N}_{t} = \boldsymbol{M}(\varepsilon_t;\boldsymbol{q}) \cdot \boldsymbol{N}_{t-1} \;,
\end{equation}
Here $\boldsymbol{M}(\varepsilon_t;\boldsymbol{q})$ is a random matrix because $\varepsilon_t$ is a random variable. The temporal sequence of $\varepsilon_t$ is randomly drawn according to the distributions of good and bad years. The long-term growth rate $\Lambda$ of the population is then given by the Lyapunov exponent of the product of these random matrices \cite{Crisanti2012}, which is calculated numerically (see methods in Appendix~\ref{sec:numerical}).

\begin{figure}
\centering
\includegraphics[width=3.4in]{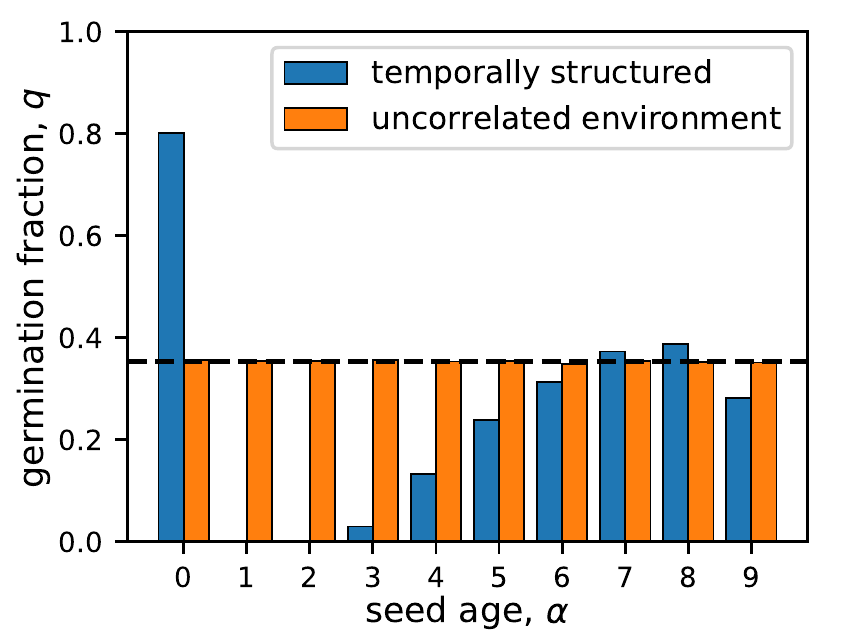}
\caption{\small Dependence of the germination fraction $q$ on the seed age $\alpha$ that maximizes the population growth rate. Blue bars are when the environment is temporally structured, as described by the duration of good and bad years in Fig.~\ref{fig:durations}B. Orange bars are when the environment is drawn independently each year, for which the germination fraction need not depend on seed age and is equal to the bet-hedging solution in Eq.~\ref{eq:Cohensol} (dashed).}
\label{fig:optimalstrategies}
\end{figure}

We vary the age-dependent germination fractions $q_\alpha$ to maximize $\Lambda$. As expected, this growth rate using seed age as an internal cue ($\Lambda_\text{int}$) is greater than that of bet-hedging without cues ($\Lambda_\text{bet}$), as illustrated in Fig.~\ref{fig:levels}B (see also Fig.~\ref{fig:memory} below). The optimal germination fraction as a function of seed age is shown in Fig.~\ref{fig:optimalstrategies}. An intuitive explanation for the age dependence is that, in this example, the bad environment typically lasts a number of years, so it is advantageous for a seed to stay dormant for a similar period of time to wait it out. Those that germinate in the wrong phase of the bad year cycle will be eliminated by selection, and the remaining individuals tend to be synchronized with the environment. In contrast, if there is no temporal structure in the environment, such as when the environment is randomly and independently chosen each year, then the seed age will no longer be correlated with the environment. In that case, the best strategy is to have a constant germination fraction (equal to $q^*$ in the bet-hedging case, see Fig.~\ref{fig:optimalstrategies}), as argued in \cite{Valleriani2006}.

Note that the information about the environment is contained in the distribution of seed ages within the population, which results from selection in previous years. Compared to the case of an external cue that is shared by all individuals, the seed age varies among individuals (which prevents an analytic expression for $\Lambda$). It acts as an individual's memory of its own lineage history, which helps it infer the likely environment in the future. Importantly, the increase in population growth rate does not come at any cost associated with sensing external cues. Thus, such an internal source of information proves to be beneficial for the population.

\subsection{Internal states as memory} \label{sec:memory}

\begin{figure}
\centering
\includegraphics[width=3.4in]{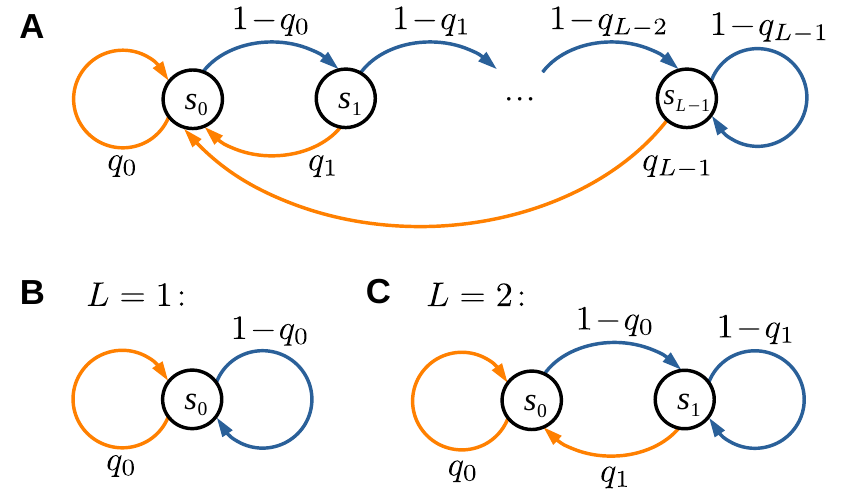}
\caption{\small State diagrams for age-dependent germination. (A) The germination fraction $q$ depends on the seed age $\alpha$ up to $\alpha = L\!-\!1$, beyond which it remains the same. Varying the length $L$ effectively varies the memory capacity of the organisms. (B) With only one state ($L=1$), the organism effectively has no memory, and the germination fraction is a constant, corresponding to simple bet-hedging. (C) The two-state case corresponds to a Markov process where the organisms switch back and forth between two phenotypes, with transition probabilities $P(\phi_1|\phi_0) = q_1$ and $P(\phi_0|\phi_1) = 1 \!-\! q_0$.} 
\label{fig:ageStructure}
\end{figure}

We have shown that internal states of organisms may help them ``remember'' the past outcomes of selection to be able to predict the future environment, leading to an increased population growth rate. Intuitively, the more the organisms can remember, the better they may predict and adapt to the environment. To test this in our model, we can vary the memory size by changing the number of possible internal states. The state diagram in Fig.~\ref{fig:durations}C has potentially an infinite number of states. They can be truncated at a finite number $L$, such that seeds exceeding age $(L-1)$ will remain in the state $s_{L-1}$ until they germinate or perish (Fig.~\ref{fig:ageStructure}A). This allows us to study how the population growth rate depends on the number of states $L$.

\begin{figure}
\centering
\includegraphics[width=3.4in]{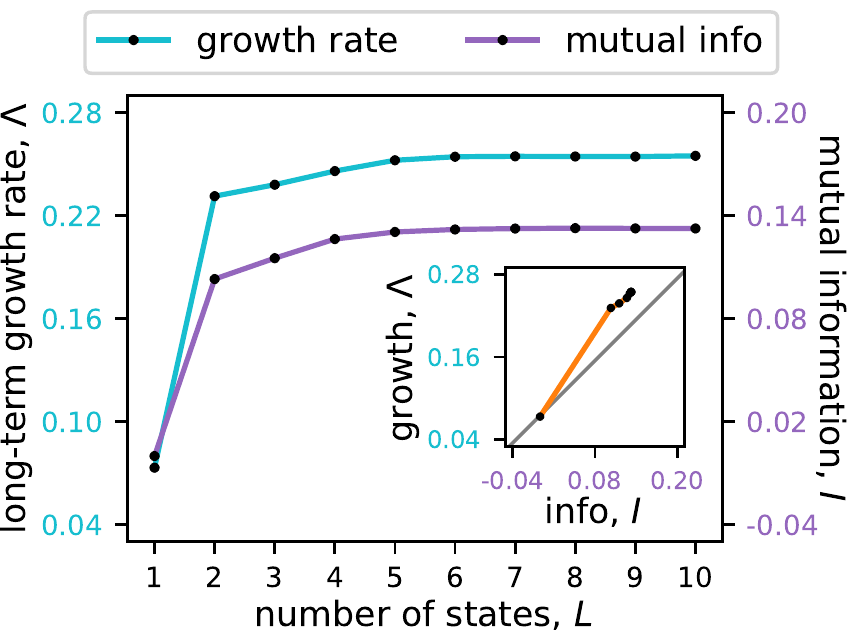}
\caption{\small Long-term growth rate $\Lambda$ of populations that have different memory capacity as measured by the number of internal states $L$. For each $L$, the age-dependent germination fractions $q_\alpha$ are chosen to maximize $\Lambda$. Also plotted is the mutual information $I$ between the previous seed age $\alpha_{t-1}$ and the environment $\varepsilon_{t}$. Both $\Lambda$ and $I$ increase monotonically with the memory capacity $L$, approaching their respective limits as $L \gg 5$ (mean duration of bad years). (Inset) Long-term growth rate $\Lambda$ increases monotonically with the mutual information $I$. Gray diagonal line represents Cohen's model with external cues, in which $\Lambda = \Lambda_\textrm{bet} + I$.}
\label{fig:memory}
\end{figure}

We first note that having only one internal state ($L=1$, Fig.~\ref{fig:ageStructure}B) is effectively having no memory, because the system will always be in that same state regardless of the past events. In this case, the germination fraction is always equal to $q_0$ associated with the only state $s_0$. Having a constant germination fraction means that this case corresponds to the simple bet-hedging strategy. The maximum long-term growth rate will just be $\Lambda_\textrm{bet}$ achieved at $q_0 = q^*$ found in Eq.~(\ref{eq:Cohensol}).

For two internal states ($L=2$, Fig.~\ref{fig:ageStructure}C), the model reduces to ``phenotypic switching'', in which the organisms randomly switch between two phenotypes (germination or dormancy) with fixed transition probabilities. Specifically, the probability for a dormant seed to germinate next year is $q_1$, and the probability for a new seed (that came from a germinated plant) to go dormant is $1 - q_0$. This is a Markov process, for which the transition between phenotypes does not depend on how long a phenotype has lasted. It implies that the germination fraction only depends on whether the seed is fresh (age 0) or has been dormant (age $>0$), but not on how long it has been dormant. As a result of being Markovian, the duration of the dormant phenotype will be geometrically distributed.

A larger $L$ will allow the germination fraction to depend more sensitively on the seed age ($L>2$, Fig.~\ref{fig:ageStructure}A). The number of states $L$ roughly represents how many dormant years a seed can remember. For each number $L$, we search for the maximum long-term growth rate $\Lambda$ over the parameters $\{ q_0, \cdots, q_{L-1} \}$ (see methods in Appendix~\ref{sec:numerical}). As shown in Fig.~\ref{fig:memory}, $\Lambda$ increases monotonically as more states are incorporated. Therefore, more memory allows faster population growth and hence better adaptation to environmental variation. Note that $\Lambda$ quickly approaches a limit $\Lambda_\textrm{mem}$ when $L$ becomes greater than the typical duration of the bad environment (equal to 5 in this example, see Fig.~\ref{fig:durations}B). Intuitively, there is no need to remember longer dormancy because there is no benefit in staying dormant for longer than the duration of bad years. The relation between the growth rate and memory is illustrated schematically in Fig.~\ref{fig:levels}B.

If we think of seed age as an internal cue for the environment, we can calculate the mutual information $I(\varepsilon_{t};\alpha_{t-1})$ between the environment $\varepsilon_t$ and the seed age $\alpha_{t-1}$, using the joint probability $P(\varepsilon_{t},\alpha_{t-1})$ calculated the same way as in Sec.~\ref{sec:internal}. Fig.~\ref{fig:memory} shows that the mutual information also increases with the number of states $L$, as more memory is available. When plotted against each other, the long-term growth rate $\Lambda$ increases with the mutual information $I$ (Fig.~\ref{fig:memory} inset), just like for an external cue. Note that in Cohen's model with external cues \cite{Cohen1967}, $\Lambda$ is simply proportional to $I$ (see Eq.~(\ref{eq:mutualinfodecom}) in Appendix~\ref{sec:derivation}). In comparison, for the same amount of information $I$, the population achieves a higher growth rate $\Lambda$ using seed age as an internal cue (Fig.~\ref{fig:memory} inset).

\begin{figure}
\centering
\includegraphics[width=3.4in]{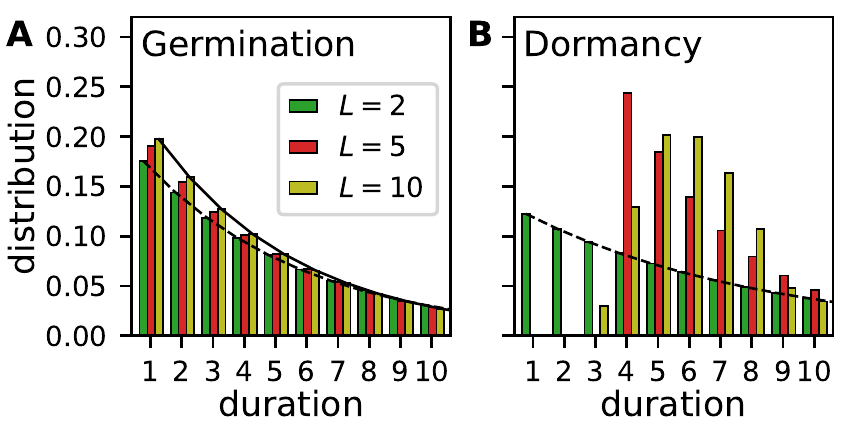}
\caption{\small The distribution of the duration of consecutive germinations or dormant years along a lineage of seeds. Different colors correspond to age-dependent germination fractions $q_\alpha$ for different memory capacities $L$. (A) For each $L$, the duration of germinations matches a geometric distribution with a mean of $1/q_0$ (dashed line for $L=2$ and solid line for $L=10$), meaning that there is no memory of previous germinations. (B) The duration of dormancy has a distribution that changes shape depending on the memory capacity $L$. $L=2$ (phenotypic switching) results in a geometric distribution with a mean of $1/(1 \!-\! q_1)$ (dashed line). Larger $L$'s result in deviation from a geometric distribution, which is indicative of having internal memory.}
\label{fig:phenotypes}
\end{figure}

So far we have considered a very specific structure for the state diagrams (Fig.~\ref{fig:ageStructure}A, ``age-diagram''). It might be possible that, given the number of internal states, there are other diagrams that can lead to a high long-term growth rate. Such diagrams could represent other types of internal states instead of the age. For example, the reserve level of an organism can be represented by a linear diagram, such that the organism moves up one or more states if it succeeds in foraging or moves down one state if it fails \cite{Higginson2018}. To find which structure of internal states provides the highest long-term growth rate for the population, we searched all possible diagrams of a given number of states (up to $L=6$, beyond which it is computationally difficult), optimizing the weights $q_\alpha$ for each diagram (see Appendix~\ref{sec:search}). It turns out that the age-diagram in Fig.~\ref{fig:ageStructure}A is optimal for the temporal structure of the environment that we assumed (Fig.~\ref{fig:durations}B). In general, the state diagram is a mathematical representation of memory, known as the ``$\epsilon$-machine'' of a stochastic process \cite{Shalizi2001}; a formal treatment and application to population growth in varying environments is given by \cite{Marzen2018}.

\section{Discussion}

\subsection{Characterization of internal memory} \label{sec:duration}

Memory arising from age-dependent germination fractions can be characterized by the distribution of the \emph{duration of dormancy}. That is, given a large number of fresh seeds, what is the distribution of the time that each seed stays dormant before germinating. To calculate this distribution, we simulate one lineage of seeds over a long time in the absence of selection (see Appendix~\ref{sec:lineage}), and record the sequence of phenotypes, i.e., whether a seed germinated or not each year. Fig.~\ref{fig:phenotypes} shows the distribution of the number of consecutive years that successive seeds germinate or that a seed stays dormant. The number of consecutive germination years is geometrically distributed with a mean of $1/q_0$ (Fig.~\ref{fig:phenotypes}A), because every new seed has the same probability $q_0$ of germinating. In other words, a new seed has no memory of the age of the plant that it came from. Thus, the absence of phenotypic memory is signified by the geometric distribution.

On the other hand, the distribution of the consecutive dormant years (i.e., the duration of dormancy) depends on the number of internal states $L$. For $L=2$, as discussed in Sec.~\ref{sec:memory}, there is no memory of how long a seed has been dormant. Indeed, the distribution of dormancy durations is geometric with a mean of $1/(1 \!-\! q_1)$ (Fig.~\ref{fig:phenotypes}B). But as $L$ increases, the distribution becomes more bell-shaped and closer to the distribution of consecutive bad years (Fig.~\ref{fig:durations}B). (In the limit where the fitness matrix $f_{\varepsilon\phi}$ is diagonal, the optimal strategy will be such that the duration of each phenotype exactly matches the distribution of the corresponding environment; see Appendix~\ref{sec:extreme}). The deviation of the distribution from being geometric indicates that the seed has memory of how long it has been dormant, which is necessary for the germination fraction to depend on the seed age. Thus, the shape of the dormancy distribution can be used as an experimental signature of internal memory.

The best demonstration of memory in phenotypic changes is found in experiments on the bacteria \textit{Bacillus subtilis} \cite{Norman2013}. During its growth, \textit{B. subtilis} can switch between two phenotypes, either as a free-moving cell by making flagela or as part of an aggregate by producing extracellular matrix \cite{Kearns2005, Lopez2009}. It is thought that the aggregate cells have an advantage for colonization and can better cope with a harsh environment by sharing resources, whereas the motile cells are better at dispersing and searching for nutrients. The durations of these two cell types along continuous cell lineages are measured in a constant environmental condition \cite{Norman2013}. It was found that the time a lineage stays in the motile cell type follows an exponential distribution with a mean of $\sim 81$ generations, while the aggregate cell type is maintained for a narrowly distributed duration with a mean and standard deviation of $7.6 \pm 2.1$ generations (see Fig.~2(d,f) of \cite{Norman2013}). This implies that the motile cell type is memoryless while the aggregate cell type has memory. That is, an aggregate cell keeps track of how long it has been part of an aggregate, whereas a motile cell turns off motility with a fixed probability at every cell division. These two distributions of phenotype durations look similar to those found in our model (Fig.~\ref{fig:phenotypes}). Importantly, since the switching of cell types is measured in a constant environment, it is evident that the phenotypic changes are influenced by some internal states of the cell, rather than external cues. This method of inferring the existence of internal memory by measuring the duration of phenotypes can be potentially applied to seeds. It would require measuring the duration of seed dormancy by planting seeds in separate pots under the same environmental condition and recording how soon they germinate.

\subsection{Evidence for age-dependent dormancy}

Our model assumes that the probability of a seed entering or exiting dormancy depends on the age. If the bad environment typically persists for a number of years, then the model predicts that the probability of exiting dormancy should be small initially and increase over a timescale that matches the duration of bad years (Fig.~\ref{fig:optimalstrategies}). Data from past experiments have shown that for different species the germination fraction can either increase or decrease between the first and second years \cite{Philippi1993}, while data going beyond the second year are scarce. To test the above prediction also requires knowing the statistics of bad years. Alternatively, age-dependent germination can be tested by measuring the distribution of dormancy durations, as discussed in Sec.~\ref{sec:duration} (Fig.~\ref{fig:phenotypes}B). For that purpose, one has to measure the final age of seeds right before they germinate. Studies on seed age structure have been done in the past \cite{Kalisz1991, Kalisz1992}, but with the goal of measuring the current age of seeds in a population at a given time, even though some seeds will continue to be dormant. We are not aware of existing studies that measured the distribution of final seed ages.

Dormancy in other organisms can also be studied using our model. One example is insect diapause \cite{Menu2002}, which is considered another example of bet-hedging. In many insect species, the larvae can enter diapause at a certain developmental stage to avoid unfavorable conditions, instead of proceeding with normal development to become adults. In a simple model of diapause \cite{Rajon2014}, the larvae may undergo multiple years of diapause and have a fixed probability of (re)entering diapause each year (see Fig.~1 of \cite{Rajon2014}), similar to Cohen's model of seed dormancy \cite{Cohen1966}. This would correspond to our model with $L=1$, such that the decision to enter diapause is memoryless. Another model assumes that the larvae can only undergo one period of diapause and must exit after that \cite{Tuljapurkar1993}. This pattern is a special case of our model with $L=2$, where the state $s_0$ would correspond to a new larva and $s_1$ to diapause. The larva can either develop to an adult with probability $q_0$ and produce offspring (arrow from $s_0$ back to itself), or enter diapause with probability $1 - q_0$ (arrow to $s_1$). However, once it undergoes diapause, it must exit and develop, so there is only one arrow leaving $s_1$, which goes to $s_0$ with probability $q_1 = 1$. In this scenario, it was found that diapause is beneficial in varying environments that are temporally correlated \cite{Tuljapurkar1993}, in agreement with our results. More generally, one may study situations where diapause can be repeated for a number of times, which would correspond to a diagram like Fig.~\ref{fig:ageStructure}A. Our results suggest that which form of diapause is evolutionarily favored depends on the complexity of temporal structure in the environmental variation, which could potentially be tested in empirical studies.

\section{Conclusion}

We have shown that the internal states of organisms can serve as a memory to help the population adapt in varying environments. In order for this strategy to be useful, the environment must be temporally structured, and the internal states must become correlated with the environment. We have demonstrated that such correlation can arise from selection alone, without direct interaction with the environment. More generally, some internal states of organisms may be correlated with the environment as a result of phenotypic plasticity. For example, seeds produced in a good year may be bigger than those produced in a bad year, so seed size could provide a memory of the past environment. It is known that seed size can affect germination probability \cite{Larios2014}, and it will be interesting to study if such dependence can benefit population growth in varying environments.

Organisms are complex systems with a lot of internal degrees of freedom, some of which might happen to become correlated with the environment through selection or plasticity. Even though these internal states might not have developed as sensors for environmental cues, they could be co-opted as information sources to guide the organism's behavior. To test whether seed age could be co-opted to affect germination, one might compare accessions of annual plants in temporally structured environments and those in unpredictable environments. Our model predicts that the germination fraction would evolve to depend on the seed age in the former case.

Dormancy has been proposed to cause a ``storage effect'' that promotes species coexistence in varying environments \cite{Chesson2000}. Our model of age-dependent dormancy may be studied in such community ecology context. If the presence of other species is viewed as part of the environment for the focal species, then internal states such as seed age could potentially provide a memory of past interaction with those other species. For example, reserve level of the predator may be an indicator of past encounters with prey \cite{Higginson2018}. History-dependent ecological interactions have been experimentally indicated in microbial communities \cite{Frentz2015}. It will be interesting to use our framework to study such ecological dynamics of organisms whose phenotypes depend on their memory.

\appendix

\section{Methods}
\setcounter{equation}{0}
\renewcommand\theequation{\thesection\arabic{equation}}

\subsection{Analytic derivation of Cohen's model} \label{sec:derivation}

Consider a population of annual plant seeds, each of which can either germinate ($\phi \!=\! 1$) or stay dormant ($\phi \!=\! 0$) each year. The environment can be either good ($\varepsilon \!=\! 1$) or bad ($\varepsilon \!=\! 0$). If a seed germinates in a good year, it will reproduce and yield $Y_1$ number of seeds; but a seed germinating in a bad year will only yield $Y_0$ seeds, with $Y_1 > Y_0$ (in the main text we set $Y_0$ to $0$ for simplicity). If a seed stays dormant, then the probability that it will remain viable is $V$. For $Y_1 > V > Y_0$, it is favorable for a seed to germinate in a good year but stay dormant in a bad year. The number of seeds at year $t$ is denoted by $N_t$ and obeys the equation:
\begin{equation} \label{eq:N}
    N_t = N_{t-1} \big[ (1-q) V + q Y_{\varepsilon_t} \big] ,
\end{equation}
where $\varepsilon_t$ is the environment in that year and $q$ is the fraction of seeds that germinates. The number of seeds at year $T$ can be calculated recursively as:
\begin{equation} \label{eq:N2}
    N_T = N_0 \prod_{t=1}^T \big[ (1-q) V + q Y_{\varepsilon_t} \big] = N_0 \big[ (1-q) V + q Y_0 \big]^{T_0} \big[ (1-q) V + q Y_1 \big]^{T_1} ,
\end{equation}
where $T_\varepsilon$ is the total number of years that the environment is $\varepsilon$. The long-term growth rate $\Lambda$ is defined as the asymptotic rate of logarithmic increase:
\begin{equation}\label{eq:logN}
\Lambda \equiv \lim_{T\to\infty} \frac{1}{T} \log \frac{N_T}{N_0} = P_0 \log \big[ (1-q) V + q Y_0 \big] + P_1 \log \big[ (1-q) V + q Y_1 \big] ,
\end{equation}
where $P_\varepsilon \equiv \lim\limits_{T\to\infty}\frac{T_\varepsilon}{T}$ is the frequency of environment $\varepsilon$. The germination fraction $q^*$ that maximizes $\Lambda$ is found by setting the derivative $\frac{\partial\Lambda}{\partial q}$ to zero, which gives (assuming $q^* > 0$):
\begin{equation} \label{eq:qstar}
    q^* = \frac{V P_1}{V - Y_0} - \frac{V P_0}{Y_1 - V} \,.
\end{equation}
And the corresponding maximum growth rate $\Lambda_\textrm{bet}$ is:
\begin{equation}
\Lambda_\textrm{bet} = P_0 \log \frac{P_0 (Y_1 - Y_0) V}{(Y_1 - V)} + P_1 \log \frac{P_1 (Y_1 - Y_0) V}{(V - Y_0)} \,.
\end{equation}

If the seeds have perfect information about the future environment, then they should all germinate in good years and stay dormant in bad years. This would result in a total population $N_T = N_0 \, V^{T_0} \, Y_1^{T_1}$ instead of Eq.~(\ref{eq:N2}), which gives the maximum possible growth rate:
\begin{equation}
\Lambda_\textrm{max} = P_0 \log V + P_1 \log Y_1 \,.
\end{equation}
The difference between $\Lambda_\textrm{max}$ and $\Lambda_\textrm{bet}$ is then given by:
\begin{equation}
\Lambda_\textrm{max} - \Lambda_\textrm{bet} = - P_0 \log \frac{P_0 (Y_1 - Y_0)}{(Y_1 - V)} - P_1 \log \frac{P_1 (Y_1 - Y_0) V}{(V - Y_0) Y_1} \,.
\end{equation}
In the limit $Y_0 \to 0$ and $Y_1 \gg V$, it simplifies to:
\begin{equation}
\Lambda_\textrm{max} - \Lambda_\textrm{bet} = - P_0 \log P_0 - P_1 \log P_1 \equiv H(\varepsilon) \,,
\end{equation}
which is the entropy of the environment.

The model above can be generalized to include an external cue $\xi$ that is correlated with the environment $\varepsilon$. Assume that, given $\xi$, the seeds will germinate with probability $P(\phi\!=\!1|\xi) \equiv q_\xi$. The total number of seeds then obeys the equation:
\begin{equation}
    N_{t} = N_{t-1} \big[ (1-q_{\xi_t}) V + q_{\xi_t} \, Y_{\varepsilon_t} \big] \,,
\end{equation}
where $\xi_t$ is the cue received in year $t$. Repeating the same procedure as above, one finds that the population after $T$ years becomes:
\begin{equation} \label{eq:Nlongcue}
    N_T = N_0 \prod_{\varepsilon,\xi} \big[ (1 - q_\xi) V + q_\xi Y_\varepsilon \big]^{T_{\varepsilon\xi}} ,
\end{equation}
where $T_{\varepsilon\xi}$ is the number of years that the environment is $\varepsilon$ while the cue is $\xi$. The long-term growth rate is then given by:
\begin{equation} \label{eq:generalGrow}
    \Lambda = \sum_{\varepsilon,\xi} P_{\varepsilon\xi} \log \big[ (1 - q_\xi) V + q_\xi Y_\varepsilon \big] ,
\end{equation}
where $P_{\varepsilon \xi} = \lim\limits_{T\to\infty}\frac{T_{\varepsilon \xi}}{T}$ is the joint probability of the environment $\varepsilon$ and the cue $\xi$. The optimal germination fraction $q_\xi^*$ that maximizes Eq.~(\ref{eq:generalGrow}) is given by (assuming $q_\xi^* > 0$):
\begin{equation} \label{eq:general}
    q_\xi^* = \frac{V P_{1|\xi}}{V - Y_0} - \frac{V P_{0|\xi}}{Y_1 - V} \,,
\end{equation}
which is the same as Eq.~(\ref{eq:qstar}) except that $P_\varepsilon$ is replaced by the conditional probability $P_{\varepsilon|\xi} = \frac{P_{\varepsilon\xi}}{P_\xi}$. The maximum growth rate achieved by using the external cue is then given by plugging Eq.~(\ref{eq:general}) into Eq.~(\ref{eq:generalGrow}), which gives:
\begin{equation}
     \Lambda_{\textrm{cue}} = \sum_{\varepsilon,\xi} P_{\varepsilon \xi} \log P_{\varepsilon|\xi} +  P_0 \log \frac{(Y_1 - Y_0) V}{(Y_1-V)} + P_1 \log \frac{(Y_1 - Y_0) V}{(V - Y_0)} \,.
\end{equation}
The difference between $\Lambda_\textrm{cue}$ and $\Lambda_\textrm{bet}$ is then:
\begin{equation} \label{eq:mutualinfodecom}
    \Lambda_\textrm{cue} - \Lambda_\textrm{bet} = \sum_{\varepsilon,\xi} P_{\varepsilon \xi} \log \frac{P_{\varepsilon|\xi}}{P_\varepsilon} \equiv I(\varepsilon;\xi) \,,
\end{equation}
which is precisely the mutual information between the environment $\varepsilon$ and the cue $\xi$.

\subsection{Numerical solution for age-dependent germination} \label{sec:numerical}

In our model where the germination fraction depends on the seed age, neither the growth rate nor the optimal germination fraction has an analytic solution. Here we describe how they are calculated numerically. Since the seeds are heterogeneous in age, the population is described by a vector $\boldsymbol{N}$ with components $N_\alpha$ that represents the number of seeds of age $\alpha$. As described in the main text, the vector $\boldsymbol{N}_t$ at year $t$ obeys the equation:
\begin{equation}
\boldsymbol{N}_{t} = \boldsymbol{M}(\varepsilon_t;\boldsymbol{q}) \cdot \boldsymbol{N}_{t-1} \;,
\end{equation}
where the matrix $\boldsymbol{M}$ depends on the current environment $\varepsilon_t$ and the germination fractions $q_\alpha \equiv P(\phi\!=\!1|\alpha)$, as given in Eq.~(\ref{eq:Mat}). Thus, the population vector after a long time $T$ is:
\begin{equation}
\boldsymbol{N}_T = \bigg( \prod_{t=1}^T \boldsymbol{M}(\varepsilon_t;\boldsymbol{q}) \bigg) \cdot \boldsymbol{N}_0 \,,
\end{equation}
and the long-term growth rate is formally given by the largest Lyapunov exponent of the product of matrices:
\begin{equation} \label{eq:Lyapunov}
\Lambda = \lim_{T\to\infty} \frac{1}{T} \log \left| \prod_{t=1}^T \boldsymbol{M}(\varepsilon_t;\boldsymbol{q}) \right| \,,
\end{equation}
where $|\cdot|$ is the matrix norm, which we choose to define as the largest eigenvalue for non-negative matrices. Compared to Cohen's model, here $\Lambda$ cannot be calculated analytically because the matrix multiplications are non-commutative. To numerically calculate $\Lambda$, we simply use the above equation with a very large $T$, as the limit is expected to converge \cite{Crisanti2012}.

We first draw a sequence of $T$ random environments as follows. Define an epoch of time $\tau_\varepsilon$ as the number of consecutive years that the environment remains to be $\varepsilon$ until it switches. The good and bad epochs are drawn from the distributions:
\begin{align}
P(\tau_1 \!=\! k) &= \frac{1}{\mu_1} \bigg( 1 - \frac{1}{\mu_1} \bigg)^{k-1} \,, \quad k = 1, 2, \cdots, \infty \label{eq:Pg} \\
P(\tau_0 \!=\! k) &= \frac{1}{Z} \exp \bigg( -\frac{(k - \mu_0)^2}{2 \sigma^2} \bigg) \,, \quad k = 1, 2, \cdots, 2\mu_0\!-\!1 . \label{eq:Pb}
\end{align}
Here $\mu_\varepsilon$ is the mean duration for the epochs, $\sigma$ characterizes the variability of the bad epochs, and $Z$ is a normalization constant. For the example used in the main text (Fig.~\ref{fig:durations}B), $\mu_1 = \mu_0 = 5$ and $\sigma = 2$. 50000 epochs are drawn for each environment, with a total length $T \approx 500000$. 

To calculate $\Lambda$, we need to calculate the product $\prod_{t=1}^T \boldsymbol{M}(\varepsilon_t;\boldsymbol{q})$. For convenience, we define $\boldsymbol{M}^{(s)} \equiv \prod_{t=1}^s \boldsymbol{M}(\varepsilon_t;\boldsymbol{q})$. Then $\boldsymbol{M}^{(T)}$ can be calculated recursively by
\begin{equation} \label{eq:recur1}
    \boldsymbol{M}^{(t)} = \boldsymbol{M}(\varepsilon_{t};\boldsymbol{q}) \cdot \boldsymbol{M}^{(t-1)},
\end{equation}
We normalize $\boldsymbol{M}^{(t)}$ at every time step by the value of its largest entry, and this normalization factor $n_t$ is stored. The Lyapunov exponent is then given by $\Lambda = \frac{1}{T} \big( \sum_{t=1}^T \log n_t + \log w \big)$, where $w$ is the largest eigenvalue of the normalized $\boldsymbol{M}^{(T)}$ (which does not matter for $\Lambda$ when $T$ is large, but matters for its derivative that we calculate below).

To find the germination fractions $q_\alpha^*$ that maximizes $\Lambda$, we use the optimization routine L-BFGS-B, which allows us to impose the constraint $0\leq q_\alpha^*\leq 1$. Besides the numerical function that calculates $\Lambda$ as described above, we also supply the Jacobian of the function, i.e., the derivative $\frac{\partial \Lambda}{\partial q_\alpha}$. This requires calculating the derivative of $\boldsymbol{M}^{(T)}$ with respect to $q_\alpha$, which can be done using the recursive relation
\begin{equation} \label{eq:recur2}
    \frac{\partial \boldsymbol{M}^{(t)}}{\partial q_\alpha} = \frac{\partial \boldsymbol{M}(\varepsilon_{t};\boldsymbol{q})}{\partial q_\alpha} \cdot \boldsymbol{M}^{(t-1)} + \boldsymbol{M}(\varepsilon_{t}; \boldsymbol{q}) \cdot \frac{\partial \boldsymbol{M}^{(t-1)}}{\partial q_\alpha} \,,
\end{equation}
together with that for $\boldsymbol{M}^{(t)}$ in Eq.~(\ref{eq:recur1}), from $t=1$ all the way to $T$. We normalize $\frac{\partial \boldsymbol{M}^{(t)}}{\partial q_\alpha}$ by the same factor $n_t$ as for $\boldsymbol{M}^{(t)}$ at every time step. The derivative of $\Lambda$ is then given by
\begin{equation}
\frac{\partial \Lambda}{\partial q_\alpha} = \frac{1}{T} \, \frac{1}{|\boldsymbol{M}^{(T)}|} \, \frac{\partial |\boldsymbol{M}^{(T)}|}{\partial q_\alpha} = \frac{1}{T} \, \frac{1}{w} \, \bigg( \boldsymbol{u} \cdot \frac{\partial \boldsymbol{M}^{(T)}}{\partial q_\alpha} \cdot \boldsymbol{v} \bigg) \,,
\end{equation}
where $\boldsymbol{u}$ and $\boldsymbol{v}$ are the left and right eigenvectors of $\boldsymbol{M}^{(T)}$ corresponding to its largest eigenvalue $w$. This derivative is then supplied as the Jacobian to the L-BFGS-B optimization routine to find the optimal $\boldsymbol{q}^*$ that maximizes $\Lambda$.

\subsection{Simulating a lineage} \label{sec:lineage}

Simulation of a continuous lineage of seeds is used to estimate the joint probability $P(\varepsilon_{t}, \alpha_{t-1})$ of the environment $\varepsilon_{t}$ and the seed age $\alpha_{t-1}$ in Sec.~\ref{sec:internal}, which is then used to calculate their mutual information $I(\varepsilon_{t}; \alpha_{t-1})$ in Sec.~\ref{sec:memory}. For a given set of germination fractions $q_\alpha$, the simulation is done as follows. We start from a fresh seed of age 0. The sequence of environments, $\{ \varepsilon_1, \cdots, \varepsilon_T \}$, is drawn beforehand as described in Sec.~\ref{sec:numerical}.

In each year, we decide whether the seed germinates or not using the germination probability that corresponds to its age. To account for selection bias, we weight the probabilities by the fitness values in the current environment. That is, in year $t$, the seed along the lineage has probability $$\frac{q_{\alpha_{t-1}} Y_{\varepsilon_t}}{q_{\alpha_{t-1}} Y_{\varepsilon_t} + (1 - q_{\alpha_{t-1}}) V}$$ to germinate and reset the age to $0$, and otherwise stays dormant with its age increased from $\alpha_{t-1}$ to $\alpha_t = \alpha_{t-1} + 1$. We repeat this procedure from $t=1$ to $T$, recording the sequence of $\alpha_t$. Afterwards, the number of times that the pair $(\varepsilon_t, \alpha_{t-1})$ takes a particular combination of values is counted, which is then normalized to be the joint probability distribution $P(\varepsilon_t, \alpha_{t-1})$, from which the mutual information $I(\varepsilon_t; \alpha_{t-1})$ is calculated.

Lineage simulation is also used to calculate the distribution of dormancy duration in Sec.~\ref{sec:duration}, i.e., the distribution of how many consecutive years a seed stays dormant in the absence of environmental variation. To calculate this distribution, we once again start with a fresh seed of age 0 and use the probability $q_0$ to decide if the seed germinates. This time the probability is not weighted by the fitness because we are calculating the dormancy durations in the absence of selection. The above procedure is repeated for a long period of time $T$ and the sequence of phenotypes at each time step is recorded as $\phi_t$. The duration of germination or dormancy is calculated by parsing the sequence of phenotypes $\{ \phi_t \}$ into consecutive epochs of germination or dormancy. The distribution of their durations is then calculated by normalizing the histograms of these epochs. Note that these distributions can also be calculated using Eq.~(\ref{eq:duration-analytic}) in Appendix~\ref{sec:extreme}.

\subsection{Exhaustive search of state diagrams} \label{sec:search}

To verify that the age-diagram in Fig.~\ref{fig:ageStructure}A is the optimal topology, we test all possible state diagrams for up to 6 internal states. For a diagram with $L$ states, we label the states as $s_0, s_1, \cdots, s_{L-1}$. Each state has two outgoing arrows, corresponding to either dormancy or germination. Each arrow can go to any other state or loop back. Therefore, naively, there can be $L^{2L}$ possible diagrams. However, many of these diagrams are equivalent in the sense that they are simply permutations of the states. To remove the redundant diagrams, we use a ``sieve'' method as follows. We first represent a diagram by a $(L \times 2)$ integer matrix, whose entry of the $\alpha$-th row and $\phi$-th column represents which state the system will transition to if it is at age $\alpha$ and expresses phenotype $\phi$. The diagrams are then indexed by a number that results from flattening the matrix and treating it as a base-$L$ number. Then, we enumerate all $L^{2L}$ diagrams starting from the index 0. For each diagram, we find all its permutations and remove their indices from the list. Furthermore, we exclude diagrams that have two or more disjoint parts to keep only connected diagrams. We go over the list of diagrams, skipping the indices that have been removed. In the end, the total number of non-degenerate diagrams for $L = 1, 2, \cdots$ is
$$n(L) = 1, 6, 52, 892, 21291, 658885, \cdots$$
which is the number of unlabeled, strongly connected, $L$-state, 2-input automata (Sequence \href{https://oeis.org/A027835}{A027835} from OLEIS). This number grows quickly and we are only able to study diagrams for up to $L=6$.

For each of the diagrams with $L$ states, we numerically find the optimal $q_\alpha$ and the maximum $\Lambda$ as in Sec.~\ref{sec:numerical}. This is computationally intensive and is done on a computer cluster. Then, among all diagrams of $L$ states, we find the optimal diagram with the largest $\Lambda$. For up to $L=6$, it turns out that the age-diagram is the optimal diagram for our model.

\subsection{Analytical results for extreme selection} \label{sec:extreme}

In the limit of extreme selection, the fitness matrix is diagonal, i.e., $f_{\varepsilon\phi} = \left( \begin{smallmatrix} V & 0 \\ 0 & Y \end{smallmatrix} \right)$. This means, hypothetically, that a seed can survive only if it germinates in a good year or stays dormant in a bad year. In this case, the long-term growth rate $\Lambda$ and the optimal germination fractions $q_\alpha^*$ have analytical solutions. Indeed, the population becomes homogeneous because, once it encounters a good year, only the seeds that germinate will survive, and subsequently the population will consist of only fresh seeds. From then on, the seed age will be synchronized with the number of consecutive bad years, and will be reset to 0 whenever there is a good year. Let $\beta_{t-1}$ be the number of consecutive bad years right before year $t$ (which is 0 if the previous year is good). It will be equal to the seed age $\alpha_{t-1}$ of the population at the beginning of year $t$. Therefore, the seed population changes over time according to:
\begin{equation}
    N_{t} = N_{t-1} (1 - q_{\beta_{t-1}}) V  \quad \textrm{or} \quad N_{t-1} \, q_{\beta_{t-1}} Y \,,
\end{equation}
depending on whether the environment $\varepsilon_t = 0$ or $1$. Over a period of time $T$, the number of seeds will be:
\begin{equation}
    N_{T} = N_{0} \prod_{\beta} \big[ (1 - q_{\beta}) V \big]^{T_{0\beta}} \big[ q_{\beta} Y \big]^{T_{1\beta}} \,,
\end{equation}
where $T_{\varepsilon\beta}$ is the number of years that the environment is $\varepsilon$ while the previous number of consecutive bad years is $\beta$. This equation has the same form as Eq.~(\ref{eq:Nlongcue}), with the external cue $\xi$ replaced by $\beta$. The long-term growth rate has the expression 
\begin{equation} \label{eq:generalGrowExtreme}
    \Lambda \equiv \lim_{T\to\infty} \frac{1}{T} \log \frac{N_T}{N_0} = \sum_\beta P_{0\beta} \log [(1-q_\beta)V] +
    \sum_\beta P_{1\beta} \log [q_\beta Y]
\end{equation}
where $P_{\varepsilon \beta} = \lim\limits_{T\to\infty}\frac{T_{\varepsilon \beta}}{T}$ is the joint probability of the environment $\varepsilon_t$ and the number of bad years $\beta_{t-1}$. Setting the derivative $\frac{\partial \Lambda}{\partial q_\alpha} = 0$, the optimal germination fractions $q_\alpha^*$ are found to be
\begin{equation} \label{envdistribution}
    q^*_\alpha = \frac{P_{1\alpha}}{P_{0\alpha} + P_{1\alpha}} \equiv P_{1|\alpha} \equiv P(\varepsilon_t \!=\! 1 | \beta_{t-1} \!=\! \alpha) \,.
\end{equation}
Here $P(\varepsilon_t \!=\! 1|\beta_{t-1} \!=\! \alpha)$ represents the conditional probability that the coming year is good, given that there has been $\alpha$ consecutive bad years. It is related to the duration distribution of bad years, $P(\tau_0)$ from Eq.~(\ref{eq:Pb}), through
\begin{equation} \label{eq:durationExtreme}
    P(\varepsilon_t \!=\! 1|\beta_{t-1}\!=\!\alpha) = \frac{P(\tau_0 \!=\! \alpha)}{P(\tau_0 \!\geq\! \alpha)}.
\end{equation}

An important consequence of this result is that, for the germination fractions $q_\alpha^*$, the dormancy duration of the seeds (as in Fig.~\ref{fig:phenotypes}B) will have the same distribution as the duration of bad years (Fig.~\ref{fig:durations}B). This is because, by definition, $q_\alpha \equiv P(\phi_t \!=\! 1|\alpha_{t-1} \!=\! \alpha)$. Let $\delta_0$ denote the duration of dormancy, then similar to Eq.~(\ref{eq:durationExtreme}), we have
\begin{equation} \label{eq:phenodistribution}
    P(\phi_t \!=\! 1 | \alpha_{t-1} \!=\! \alpha) = \frac{P(\delta_0 \!=\! \alpha)}{P(\delta_0 \!\geq\! \alpha)} \,.
\end{equation}
Equating the left-hand sides of Eqs.~(\ref{eq:durationExtreme}) and (\ref{eq:phenodistribution}) leads to, as stated above,
\begin{equation}
P(\delta_0 \!=\! \alpha) = P(\tau_0 \!=\! \alpha) \,.
\end{equation}
Incidentally, for a general $q_\alpha$, it can be shown that
\begin{equation} \label{eq:duration-analytic}
P(\delta_0 \!=\! \alpha) = q_\alpha \prod_{k=1}^{\alpha-1} (1 - q_k) \,.
\end{equation}


\end{document}